\documentstyle[aps,twocolumn,prb,psfig]{revtex}
\newcommand{\bge}{\begin{equation}}
\newcommand{\ege}{\end{equation}}
\newcommand{\bga}{\begin{eqnarray}}
\newcommand{\ega}{\end{eqnarray}}
\newcommand{\nnu}{\nonumber}

\input epsf
\begin{document}
\draft

\title{Time Reversal Symmetry breaking superconductivity}

\twocolumn[ 
\hsize\textwidth\columnwidth\hsize\csname@twocolumnfalse\endcsname 

\author{Haranath Ghosh}
\address{
Instituto de F\'\i sica,
         Universidade Federal Fluminense,
         24210-340 Niter\'oi, Rio de Janeiro, Brazil}
\date{\today}
\maketitle

\begin{abstract}
We study time reversal symmetry breaking superconductivity
with $\Delta_k = \Delta_{x^2-y^2} (k)
+e^{i\theta} \Delta_{\alpha}$ ($\alpha = s$ or $d_{xy}$)  symmetries.
It is shown that the behavior of such superconductors
could be {\em qualitatively} different depending on 
the minor components ($\alpha$) and its phase at lower temperatures. 
It is argued that such {\em qualitatively different} 
behaviors in thermal as well as in angular dependencies 
could be a {\em source} of
consequences in transport and Josephson physics.   
Orthorhombicity is found to be a strong mechanism for mixed phase (in case of
$\alpha = s$). We show that due to electron correlation 
the order parameter is more like a pure 
$d_{x^2-y^2}$ symmetry near optimum doping.
\end{abstract}
]
%%%%%%%%%%%%%%%%%%%%%%%%%%%%%%%%%%%%%%%%%%%%%%%%%%%%%%%%%%%%%%%%%%%%%%%%%%%%%%%
During last several years, a great deal of effort has been devoted to
determine the symmetry of the superconducting order parameter (SCOP) of 
high $T_c$ cuprate superconductors \cite{1}. While there is a strong concensus
both from experiments and theory that the symmetry of the SCOP in the cuprates
could be $d_{x^2-y^2}$
type; recent studies provide increasing evidence that the pairing
symmetry may
be an admixture with a minor component, like $s, ~d_{xy}$ with predominant 
$d_{x^2-y^2}$ symmetry. The pairing symmetry provides clues to identify
pairing mechanism which is essential for the development of the theory of 
high temperature superconductivity in cuprates, still remains
a challenge after a decade of its discovery.

 Several different types of measurements which are sensitive to the phase of
$\Delta (k)$ indicate a significant mixing of $s$-wave component with
$d_{x^2-y^2}$ symmetry. For example, c-axis Josephson tunnelling studies on
junctions consisting of a conventional $s$ wave 
superconductor (Pb) and twinned or 
untwinned single crystals of YBCO indicate that the SCOP of YBCO has 
mixed $d\pm s$ (or $d\pm$ i$s$) symmetry \cite{2,3}. Recently, a new class of c-axis Josephson 
tunneling experiments are reported by Kouznetsov {\it et al}, \cite{4}
in which a conventional superconductor (Pb) was deposited
across a single twin boundary of a YBCO single crystal. There measurements of
critical current as a function of the magnitude and angle of a magnetic field
applied in the plane of the junction provides a direct evidence for mixed
$d-$ and $s$-wave pairing in $YBCO$. A series of high resolution measurements 
on thermal conductivity ($\kappa$) in the
$Bi_2Sr_2CaCu_2O_8$ by Krishna {\it et al}, \cite{5} show that the
$\kappa$ at low temperature becomes field-independent above a
temperature dependent kink field $H_k(T)$.
This remarkable result
indicates a phase transition separating a low-field state where the thermal
conductivity decreases with increasing field and a high-field one where it
is insensitive to applied magnetic field. The authors argue that this phase
transition is not related to the vortex lattice because of the temperature
dependence of the field $H_k (T)$ (which is roughly proportional to $T^2$) as
well as its magnitude. Instead, they suggest a field-induced electronic
transition leading to a sudden vanishing of the quasi-particle contribution
to the heat current. Possible scenarios would be the induction of 
a minor $e^{i\theta} \Delta_{\alpha}$ ($\alpha = d_{xy}$ or $s$)
 component with $d_{x^2-y^2}$ symmetry
 with application of a weak field.
A similar conclusion may also be obtained based on the results of
angle resolve phtoemission specroscopy 
(ARPES) experiment by J. Ma {\it et al}., \cite{6} in which a temperature
dependent gap anisotropy in
the oxygen-annealed $Bi_2Sr_2CaCuO_{8+x}$ compound was found. The measured
gaps along both high symmetry directions ($\Gamma -M$ {\it i.e}, Cu-O bond
direction in real space and $\Gamma - X$ {\it i.e}, diagonal to Cu-O bond)
are non-zero at lower temperatures and their ratio is
strongly temperature dependent. This observed feature cannot be
explained within a simple $d$-wave scenario (but is consistent with 
$d_{x^2-y^2} + e^{i\theta}\alpha$ scenario) and has been
taken as a signature of a
two component order parameter, $d_{x^2-y^2}$ type close to $T_c$ and a mixture
of both $s$ and $d$ otherwise \cite{7}. However, surprisingly enough, the 
experimental results \cite{6} was never been reproduced by any other 
equivalent experiments. In Josephson physics, it was found 
\cite{8} that the current-phase relationship depends on the mutual
orientation of the two coupled superconductors and their interface (the
induced minor component then would be associated with any arbritrary phase,
$\theta$). This
property is the basis of all the phase sensitive experiments probing
SCOP. Therefore, motivated by strong experimental evidences of a mixed order 
parameter symmetry in high $T_c$ cuprates we present basic behaviors 
of superconductors with mixed order parameter symmetries such as 
$d + e^{i\theta} \alpha$, $\alpha$ = $s, ~d_{xy}$, which is characterized
by a local breakdown of time reversal symmetry. 
A clear evidence will emerge that the time reversal superconductivity at lower
temperatures give rise to {\em unusual} behaviors depending on $\alpha$ and
its phase.

% The high $T_c$ cuprates are known to be strongly correlated systems and 
%a quasiparticle picture may not be valid in the normal state. 
Assuming a well defined quasiparticle picture
in the superconducting (SC) state, free energy of a
superconductor with arbritary pairing symmetry may be written as,
\begin{equation}
F_{k,k^\prime} = -\frac{1}{\beta} \sum_{k,p = \pm} \ln (1 + e^{-p \beta E_k}) +
\frac{\mid \Delta_k \mid^2}{V_{k k^\prime}}
\end{equation}
where $E_k =  \sqrt{(\epsilon_{k}-\mu)^2 + \mid \Delta_k \mid^2}$ 
are the energy 
eigen values of a Hamiltonian that describes superconductivity. It is generally
believed that a single band Hubbard model contains essential ingredients to
describe $Cu-O$ planes where transport processes takes place. Therefore, our 
model Hamiltonian reads as,
\begin{eqnarray}
& &{\cal H}  =  H_{HUB} + H_{PAIR}  \\
&&
H_{HUB} = \sum_{ij\sigma} t_{ij} c_{i\sigma}^\dagger c_{j\sigma} + U\sum_i
n_{i \uparrow}n_{i\downarrow}
- \mu \sum_{i\sigma} n_{i\sigma} \\ 
&& H_{PAIR} = \sum_{kk^\prime}V_{kk^\prime}
c_{k\uparrow}^\dagger c_{-k\downarrow}^\dagger c_{-k^\prime \downarrow}
c_{k^\prime \uparrow}
\end{eqnarray} 
where the notations have their usual meanings.

Treatment of the Hubbard on-site correlation is a known problem.
We use the
Slave Boson theory of Kotliar and Ruckenstein \cite{9}in which
the Hubbard model has solution for any filling and any value of repulsion.
The slave - boson formulation of Kotliar and
Ruckenstein (KR) \cite{9} introduces four auxilliary boson
fields corresponding to the occupancy of a site, namely empty
($e_i$), doubly occupied ($d_i$), singly occupied ($p_{i\sigma}$/
$p_{i-\sigma}$) with spin $\pm \sigma$ respectively. In terms
of the slave boson field operators the single band Hubbard model
takes the form
\bga
H_{HUB} & = &\sum _{ij, \sigma} t_{ij} z_{i
\sigma}^{\dagger}c_{i\sigma}^{\dagger} c_{j\sigma}z_{j\sigma} +
U\sum_{i}d_{i}^{\dagger}d_i -
\mu\sum_{i\sigma}c_{i\sigma}^{\dagger}c_{i\sigma}\nnu \\
&& +
\sum_{i}\lambda_{i}(1-e_i^{\dagger}e_{i} 
- d_{i}^{\dagger}d_{i}-\sum_{\sigma}
p_{i\sigma}^{\dagger}p_{i\sigma}) \nnu \\
&& +
\sum_{i\sigma}\lambda^{\prime}_{i
\sigma}(c_{i\sigma}^{\dagger}c_{i\sigma}
-d_{i}^{\dagger}d_i-p_{i\sigma}^{\dagger}p_{i\sigma})
\ega
where
$z_{i\sigma}=(1-d_i^{\dagger}d_i-p_{i\sigma}^{\dagger}p_{i\sigma})^{-1/2}
(e_i^{\dagger}p_{i\sigma}+p_{i-\sigma}^{\dagger}d_i)(1-e_{i}^{\dagger}e_i
-p_{i-\sigma}^{\dagger}p_{i-\sigma})^{-1/2}$
and the Boson field operators are not independent of each other
but are constrained by the requirements of completeness and
local charge conservation at a site, hence the fourth and fifth
terms are added in the Hamiltonian with the help of Lagrange
multipliers ($\lambda_i$ and $\lambda^{\prime}_i$).
The values of the boson field operators and Lagrange multipliers
are determined by minimizing the free energy in the saddle point
approximation. This approximation diagonalizes the $U$ term
whereas renormalizes the hopping term as $\tilde q t_{ij}$ with
$\tilde q = z^{\dagger}z$ ; $\tilde q$ usually being complicated
function of $u$ ($={U\over U_c}$), dopant concentration ($\delta
$) and $x (=e+d)$ carries the necessary information of electron
correlation. In this approach solutions are obtained for the
paramagnatic state for all values of $u$ and band fillings
\cite{7} that reproduces the correct Brinkman-Rice result for
metal-insulator transition ($U_c$) at half-filling. 
 
 As a consequence the total Hamiltonian (2) takes the form,
\bga
{\cal H}& =&\sum_{k\sigma}(\tilde q \epsilon_k - {\mu}) c_{k\sigma}^\dagger
 c_{k\sigma} + \sum_{kk^\prime}\tilde q ^2 V_{kk^\prime}
c_{k\uparrow}^\dagger c_{-k\downarrow}^\dagger c_{-k^\prime \downarrow}
c_{k^\prime \uparrow} \nnu \\
&& +~bosonic ~terms
\ega
where the band dispersion  $\epsilon_k = \sum_{i=1}^5 C_i \eta_i (k)$ 
are obtained from the
angle resolved photoemission spectroscopy (ARPES) results \cite{10} on 
$Bi_2Sr_2CaCuO_8$. In deriving Hamiltonian (6) fluctuations of the bose-field
operators are {\it not} considered (consequence of on-site correlation has 
shown up in (6) as $\epsilon_k \to  \tilde q \epsilon_k$ and 
$V_{kk^\prime} \to \tilde V_{kk^\prime} = \tilde q ^2 V_{kk^\prime}$).

A mean field superconducting (SC) order parameter
 $\Delta^\star (k) = 
\sum_{k^\prime} \tilde V_{kk^\prime} <c_{k^\prime \uparrow}^\dagger 
c_{-k^\prime \downarrow}^\dagger > $ allows to solve the Hamiltonian (6)
 exactly to find 
the energy eigen values $E_k$ and hence minimization of the free energy (1)
(i.e $\partial F_{k,k^\prime} / \partial \Delta_k = 0 $) 
will result the superconducting (SC) gap equation as,
\begin{equation}
\Delta_k = \sum_{k^\prime}\tilde V_{kk^\prime} \frac{\Delta_{k^\prime}}
{2 E_{k^\prime}}
\tanh (\frac{\beta E_{k^\prime}}{2}),
\end{equation}
\noindent where the pairing potential $V_{kk^\prime}$ for a two component order
parameter with a separable form and the corresponding gap functions
are obtained as,
\begin{equation}
V_{kk^\prime}=\sum_{j=0,}^1 V_j f_{k}^j f_{k^\prime}^j~~~ \& ~~
\Delta_k = \sum_{j=0,}^1 e^{j(i\theta)} \Delta_j f_{k}^j
\end{equation}
\noindent For a mixed $d_{x^2-y^2} - d_{xy}$ symmetry, $V_{0(1)}$ 
$\equiv$ $V_{d_{x^2-y^2}}$ 
$(V_{d_{xy}}$) and $f_{k}^{0(1)}\equiv \cos k_x - \beta^\prime \cos k_y 
(2 \sin k_x \sin k_y)$
and the corresponding gap amplitudes are 
$\Delta_{0(1)} \equiv \Delta_{d_{x^2-y^2}} (\Delta_{d_{xy}})$
($\beta^\prime \le 1$ corresponds to orthogonal or tetragonal symmetry). 
Similarly for a 
mixed  $d_{x^2-y^2} - s$ symmetry, $V_{1} \equiv V_s, 
~f_{k}^1\equiv$ constant and $\Delta_{1} \equiv \Delta_s$.
Substituting equation (8) in (7) the gap equations for different 
components can be separated out as,
\begin{equation}
\Delta_j = \sum_{k} V_{j} \frac{\Delta_j {f_{k}^j}^2}{2 E_k}
\tanh (\frac{\beta E_{k}}{2})
\end{equation}
The two coupled gap equations (9), a number conserving equation to fix chemical 
potential ($\mu$) and a sixth order algebric equation in $x (= e+d)$
(to get correct values of $\tilde q$ for any value of $u,~\delta$) are solved 
self-consistently for a fixed cut-off parameter $\Omega_c = 500$ K around
the Fermi level beyond which SC pairing does not exist.

In figures 1(a,b) phase diagrams of superconductors with 
$\Delta_k = \Delta_{x^2-y^2} (k)+e^{i\theta} \alpha$ where 
$\alpha = \Delta_{xy}(k)$ {\bf (a)} and =$\Delta_s$ {\bf (b)} are obtained
for a fixed band filling $\rho = 0.80$ and in a tetragonal square lattice.
The phase diagrams comprises the amplitudes of pair condensates, 
evaluated at 5 K, in different channels as a function of the ratio $\gamma 
=V_{d_{x^2-y^2}}/V_{\alpha}$. The minor components $\alpha$ appear at the
same value of $\gamma$ irrespective of the phase $\theta$ in the predominant 
$d_{x^2-y^2}$ phase. But the rate of growth of the minor component is different
depending on $\theta$.  For $\theta = 0$ {\it i.e} for a real 
order parameter with (a) $d_{x^2-y^2}+d_{xy}$ or (b) $d_{x^2-y^2}+s$ wave 
symmetry, both the minor as well as the $d_{x^2-y^2}$ component increases with
decrease in $\gamma$, thereby inducing each other. 
In case of $d +s$ symmetry this 
phenomena is very pronunced. However, for any finite $\theta (\neq 0)$ the 
$d_{x^2-y^2}$ component is suppressed with the enhancement in the 
isotropic $s$ component at some values of $\gamma$ depending on 
$\theta$ resulting in a pure $s$-wave phase. This
feature is absent in $\Delta_k = \Delta_{x^2-y^2} (k)+
e^{i\theta} \Delta_{xy}(k)$ scenario. 
%%%%%%%%%%%%%%%%%%%%%%%%%%%%%%%%%%%%%%%%%%%%%%%%%%%%%%%%%%%%%%%%%%%%%%%%%%%%%%%
\begin{figure}
\epsfxsize=5.0truein%10cm
\epsfysize=4.40truein%10cm
%\begin{center}
%\leavevmode
{\epsffile{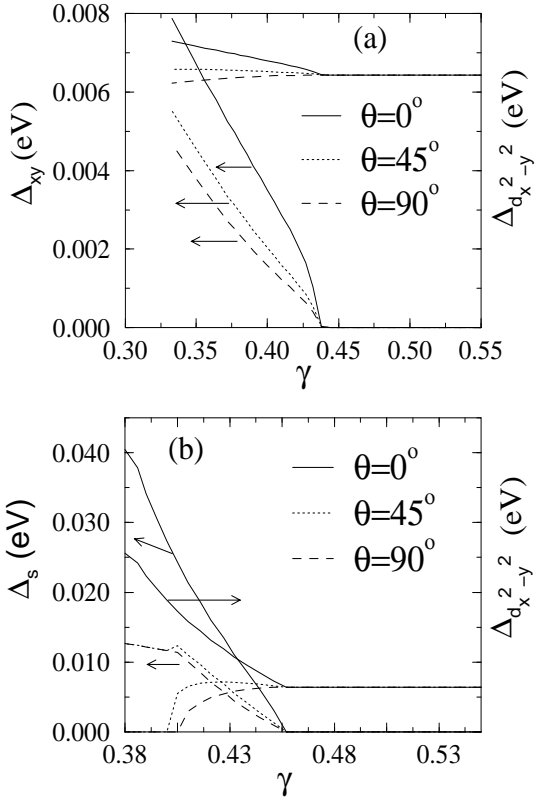}}
\caption{Phase diagrams of superconductors with 
{\bf (a)} $\Delta_k$ =$ \Delta_{x^2-y^2}$ (k)
+$e^{i\theta}$$ \Delta_{xy}(k)$ and {\bf (b)} $\Delta_k$ = $\Delta_{x^2-y^2} (k)
$+$e^{i\theta}$$ \Delta_{s}$ symmetries.
Amplitudes of the $d_{x^2-y^2}$,  $d_{xy}-$ and $s-$wave 
pairing condensates (at 5 K)
are plotted as a function of the
ratio of pairing strengths in the respective channels,
$\gamma = V_{d_{x^2-y^2}}/V_{d_{xy}}$ for $u = 0.4$.
Note the {\it qualitative difference}
in {\bf (a)} $\&$ {\bf (b)}; the onset of $s$-wave component influences
the $d_{x^2-y^2}$-amplitude very strongly (suppresses or enhances depending on
the phase $\theta$) whereas the onset of $d_{xy}$ component have no influence
on the $d_{x^2-y^2}$-amplitude except for $d_{x^2-y^2}+ d_{xy}$ phase.}
%\label{sq}
%\end{center}
\end{figure}
%%%%%%%%%%%%%%%%%%%%%%%%%%%%%%%%%%%%%%%%%%%%% 
For $\pi/2 < \theta < 0$, the minor component has two parts, real and complex (thereby giving 
rise to most probable mixing), that
couples to the $d_{x^2-y^2}$. A close look to figure 1 therefore, would
suggest that the mixing of a minor real $s$ or $d_{xy}$ component 
will stabilize
the $d_{x^2-y^2}$ symmetry, whereas inclusion of complex minor component
$s$-wave tends to suppress $d_{x^2-y^2}$ gap. The $d_{xy}$ minor component
with a complex phase has hardly any influence on the $d_{x^2-y^2}$ symmetry.
These {\it qualitatively different} behaviors in different classes of
mixed order parameter symmetries are also responsible for so with respect to
thermal dependency.
% and is thus {\it obvious source} of interesting 
%thermodynamical and transport consequences. 

It is evident from figures 2(a,b) (for $\rho = 0.8$)
that at lower temperatures the minor 
component plays a significant role to the predominant $d$-wave. For example,
when the minor component mixes with the $d$-wave 
with a real coefficient ({\it i.e,} $\theta = 0$),
the minor component has a very fast growth with lowering in temperature
and it always {\it induces} the $d$-wave component by 
further stabilizing it.
On the other hand, mixing of the minor 
component with a complex coefficient ({\it i.e,} $\theta = 90^o$) suppresses the
$d$-wave component when the minor component is $s$-wave and has no effect when
the same is $d_{xy}$. These behaviors are 
{\it bound} to have immense influence on 
thermodynamic quantities which are sensitive to the thermal gradient of the 
SC-gap. For instance, the specific heat will then have two jumps, at $T_c$
and at $T_{c}^\alpha <T_c$ where the minor component ($\alpha$) 
vanishes and the 
nature of the {\it second} jump can be very different depending on the phase 
$\theta$ providing signature of a true {\it second} transition below
$T_c$ \cite{11}.
%%%%%%%%%%%%%%%%%%%%%%%%%%%%%%%%%%%%%%%%%%%%%%%%%%%%%%%%%%%%%%%%%%%%%%%%
\begin{figure}
\epsfxsize=5.0truein%10cm
\epsfysize=4.40truein%10cm
{\epsffile{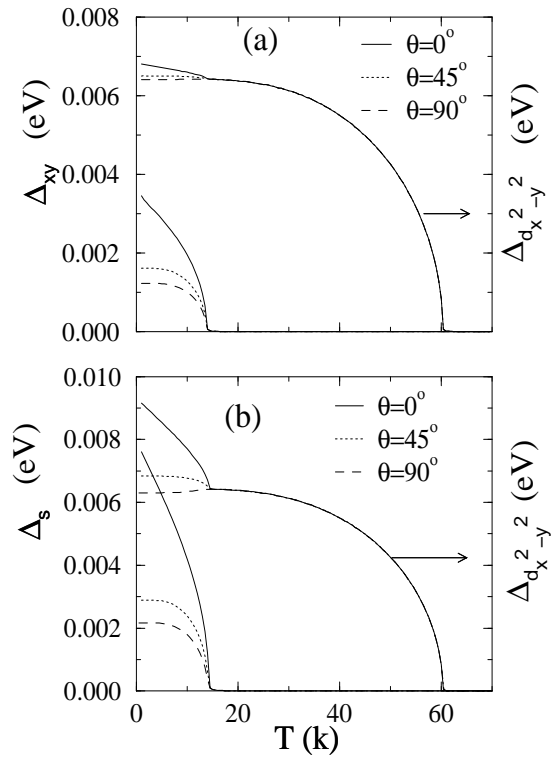}}
\caption{Thermal behavior of superconductors with 
{\bf (a)} $\Delta_k$ =$ \Delta_{x^2-y^2} (k)
$+$e^{i\theta}$$\Delta_{xy}(k)$ ($\gamma = 0.195$) and 
{\bf (b)} $\Delta_k$ = $\Delta_{x^2-y^2} (k)$
+$e^{i\theta}$$\Delta_{s}$ ($\gamma = 0.18$) 
symmetries for various values of $\theta$ ($u = 0.4$) . 
The minor ($s$ or $d_{xy}$) component although have {\em same} $T_c$ their 
thermal growth inside the $d$- component is {\em qualitatively} different. 
Consequently, the predominant $d$ component is influenced (e.g, further
stabilised for $\theta = 0$ and suppressed for $\theta = 90$) 
very {\em differently}.}
\end{figure}
%%%%%%%%%%%%%%%%%%%%%%%%%%%%%%%%%%%%%%%%%%%%%%%%%%%%%%%%%%%%%%%%%%%%%%%%
 In figures 3(a,b) we present the angular correlation between the amplitudes
of different channels evaluated at 5 K and $\rho = 0.8$. 
%It is always found that the mixing of the minor
%component is minimum for $\theta = \pi/2$ (which results in a {\it global} 
%minima in the free enery also). Note, the amplitudes of the different symmetry
%wave order parameters are derived from free energy minimization already
%discussed above.  
For any finite mixing of the minor component
with $d_{x^2-y^2}$-wave thus produces states that spontaneously
 breaks time reversal 
symmetry. Further interesting point to note is that with small orthorhombicity 
 $\beta^\prime=0.98$,(in the lattice as well), 
the $s$  component gets enhanced largely
(curves with solid circle symbol) whereas the change in  
$d$-component with orthorhombicity is 
rather dependent on $\theta$. 
For example, the predominant $d$-component increases with orthorhombocity 
for $0<\theta<0.22\pi$ or $1.78\pi<\theta<2\pi$ and decreases otherwise 
(both cases being small, cf. Fig 3(a)).  The corresponding changes in 
$\Delta_{d_{xy}}$ and associated change in $\Delta_{d_{x^2-y^2}}$ are 
negligibly small (cf. Fig. 3(b)) for such orthorhombicity. Therefore, 
orthorhombocity could be an intrinsic mechanism for $d\&s$ mixing (with
an arbritary phase between them), $d+s$ state being most favorable.

In figures 4(a,b) we demonstrate the role of electron correlation on mixed
pairing symmetry.
% (so far in figures 1--3, the value of
%on-site correlation $u(=U/U_c$) was kept fixed ($u=0.4$)).
It is shown that the minor component is always minimum at
the optimum doping and is suppressed largely due to electron correlation 
explaining the novel features observed in ARPES experiment \cite{12}. This
is based on the study of the amplitudes of the different components for the
case of $\theta = \pi/2$ ({\it i.e} $d$+i$s$ state) and for fixed
$T_{c}^\alpha$ = 45K at $u =0$ as a function of density $\rho$ (=1-$\delta$).
The amplitudes are normalized with respect to the
value of the $d$-component at optimum doping (the $d$-wave also gets
suppressed with $u$ but at a slower rate).  Note, the electronic
correlation suppresses the minor component very largely close to half-filling
%%%%%%%%%%%%%%%%%%%%%%%%%%%%%%%%%%%%%%%%%%%%%%%%%%%%%%%%%%%%%%%%%%%%%%%%%%%%
\begin{figure}
\epsfxsize=2.05truein%10cm
\epsfysize=2.95truein%10cm
\centerline{\epsffile{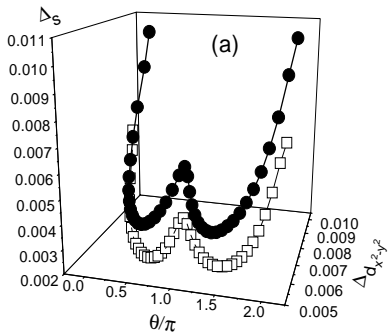}}
\epsfxsize=2.05truein%10cm
\epsfysize=2.95truein%10cm
\centerline{\epsffile{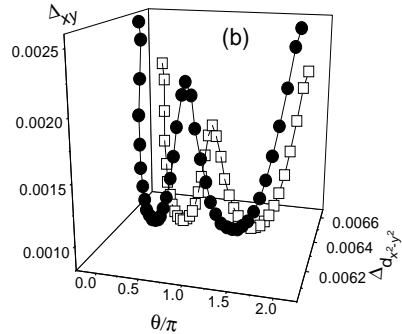}}
\caption{Amplitudes of the $d_{x^2-y^2}$ and the minor  gap {\bf (a)} $s$ or
{\bf (b)} $d_{xy}$ as a function of the phase of the minor component 
(for $u=0.4 ~\& ~T_{c}^\alpha = 14 K$). The curves with solid circles
represent orthorhombic phase with $\beta^\prime = 0.98$ whereas the curves 
with open squares correspond to tetragonal lattice.}
\end{figure}
%%%%%%%%%%%%%%%%%%%%%%%%%%%%%%%%%%%%%%%%%%%%%%%%%%%%%%%%%%%%%%%%%%%%%%%%%%%
 
\noindent whereas hardly have any effect at lower densities (the $d$-wave boundary 
shrinks a bit). 
Also, this behavior with density is similar
for any other arbritrary value of $\theta$ \cite{11}. Apparently, therefore,
close to the  optimium doping the order parameter will be
more like a pure $d_{x^2-y^2}$ like. Furthermore, no mixing is 
found within a wide range of densities between $d$-wave with other 
probable minor components like $s_{x^2+y^2}, s_{xy}$ ({\em e.g,} these 
components have maxima at $\rho \sim 0$)

 Therefore, we have presented basic consequences of
time reversal symmetry breaking superconductivity. It is demonstrated that
mixing of a minor component at lower temperatures can give rise to
{\em unusual} thermal behaviors depending on the nature of the minor component
as well as the phase associated with it. In the {\em fully gapped} phase at
lower temperatures, as described by the $\Delta_k = \Delta_{x^2-y^2} (k) +
e^{i\theta}\Delta_{\alpha}$ scenario, the number of quasiparticles will be
exponentially small and hence will contribute negligibly small to thermal
conductivity (and in any other
transport properties). This scenario therefore can
justify the experimental observation by Krishna {\em et al.,} \cite{5,7}.
This scenario also has strong potential to have natural explanation to the
experimental observation \cite{4} as discussed earlier, because mixing of
a minor component ($s$ or $d_{xy}$) with predominant $d_{x^2-y^2}$ wave would
change sign across the twin boundary (hence the situation is described in
figures 3) yielding a dramatically different angular dependence.

%%%%%%%%%%%%%%%%%%%%%%%%%%%%%%%%%%%%%%%%%%%%%%%%%%%%%%%%%%%%%%%%%%%%%%%%%%%%
\begin{figure}
\epsfxsize=4.5truein%10cm
\epsfysize=4.40truein%10cm
{\epsffile{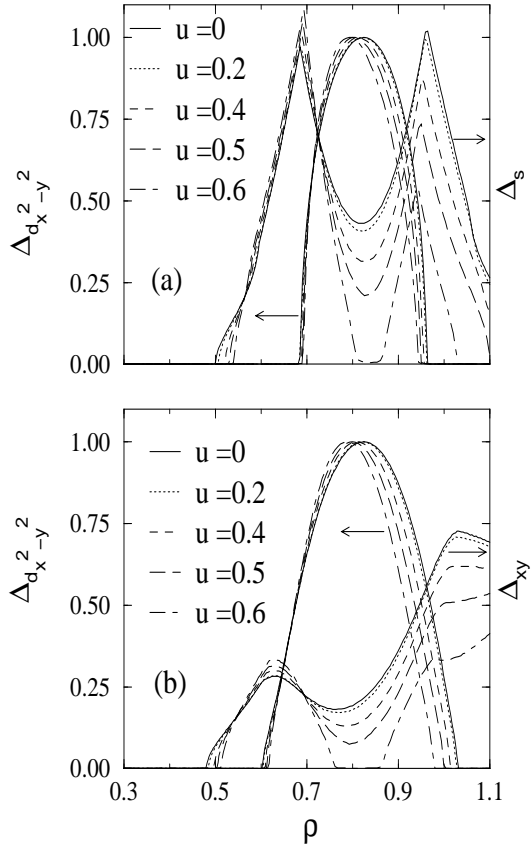}}
\caption{Amplitudes of different components of
{\bf (a)} $\Delta_k = \Delta_{x^2-y^2} (k)
+i\Delta_{s}$ and
{\bf (b)} $\Delta_k = \Delta_{x^2-y^2} (k)
+i \Delta_{xy}$ superconductors at 5K. The minor
($s$ or $d_{xy}$) component at $u=0$ has $T_{c}^\alpha (0) = 45 K$. 
The electron correlation minimizes the mixing of the minor component
at optimum doping.}
\end{figure}
%\acknowledgments
%%%%%%%%%%%%%%%%%%%%%%%%%%%%%%%%%%%%%%%%%%%%%%%%%%%%%%%%%%%%%%%%%%%%%%%%%%%%

% Therefore, we have presented basic consequences of 
%time reversal symmetry breaking superconductivity. It is demonstrated that 
%mixing of a minor component at lower temperatures can give rise to 
%{\em unusual} thermal behaviors depending on the nature of the minor component
%as well as the phase associated with it. In the {\em fully gapped} phase at 
%lower temperatures, as described by the $\Delta_k = \Delta_{x^2-y^2} (k) +
%e^{i\theta}\Delta_{\alpha}$ scenario, the number of quasiparticles will be
%exponentially small and hence will contribute negligibly small to thermal
%conductivity (and in any other 
%transport properties). This scenario therefore can
%justify the experimental observation by Krishna {\em et al.,} \cite{5,7}.
%This scenario also has strong potential to have natural explanation to the
%experimental observation \cite{4} as discussed earlier, because mixing of
%a minor component ($s$ or $d_{xy}$) with predominant $d_{x^2-y^2}$ wave would
%change sign across the twin boundary (hence the situation is described in 
%figures 3) yielding a dramatically different angular dependence. 
In particular, this study provides a significant understanding on the nature
of thermodynamic transition to a gapped or coherent phase from an ungapped or
incoherent phase. We also pointed out the subtle role of electronic correlation
with number density prohibiting mixing of the minor component at optimum 
doping.

The work is supported by the Brazilian Funding Agency FAPERJ, project no.
E-26/150.925/96-BOLSA. I thank M. Mitra and S. N. Behera for their 
contributions in using slave boson technique in the work.


\begin{references}

\bibitem{1} D. L. Cox and M. B. Maple, Physics Today, February (1995) 32 ; B. G. Levi, Physics Today, May (1993), 17 ; January (1996) 19.

\bibitem{2} A. G. Sun, {\it et al.,}
% D. A. Gajewski, M. B. Maple, and R. C. Dynes, 
Phys. Rev. Lett. {\bf 72} 2267 (1994).

\bibitem{3} A. G. Sun, {\it et al.,}
%S. H. Han,A. S. Katz, D. A. Gajewski, M. B. Maple, and R. C. Dynes, 
Phys. Rev. B {\bf 52} R15731 (1995).

\bibitem{4} K. A. Kouznetsov, {\it et al.,}
% A. G. Sun, B. Chen, A. S. Katz et al., 
\prl {\bf 79}, 3050 (1997).

\bibitem{5} K. Krishana, {\it et al.,}
%N. P. Ong,  Q. Li, G. D. Gu, N. Koshizuka, 
Science {\bf 277}, 83 (1997).

\bibitem{6} Jian Ma, {\it et al.,}
%C. Quintmann, R. J. Kelley, H. Berger, G. Margaritondo, M. Onellion, 
Science {\bf 267}, 862 (1995).

\bibitem{7} M. Mitra, Haranath Ghosh and S. N. Behera, Euro. J. Phys. B {\bf 2},
371 (1998) ; Haranath Ghosh, Europhysics Letters {\bf 43}, 707(1998).

\bibitem{8} S. Yip, \prb  {52}, 3087 (1995).

\bibitem{9} G. Kotliar, {\it et al.,} \prl {\bf 57}, 1362 (1986). 
%and A. E. Ruckenstein, \prl {\bf 57}, 1362 (1986).

\bibitem{10} M. R. Norman, {\it et al.,}
%M. Randeria, H. Ding and J. C. Campuzano,
 \prb {\bf 52}, 615 (1995).

\bibitem{11} Haranath Ghosh, submitted for publication (1998).
\bibitem{12} R. J. Kelley {\it et al.}, Science {\bf 271},1255 (1996).
\bibitem{13} For some other original works in the field see, M. Sigrist
{\it et al.} Rev. Mod. Phys. {\bf 63}, 239 (1991); M. Prohammer {\it et al.},
\prb {\bf 47}, 15152 (1993); R. Movshovich {\it et al.}, \prl {\bf 80}, 1968
(1998); A. V. Balatsky, \prl {\bf 80}, 1972 (1998); G. M. Luke {\it et al.},
Nature {\bf 394}, 558 (1998).

\end{references}
\end{document}